
%
\catcode`@=11 
\def\space@ver#1{\let\@sf=\empty \ifmmode #1\else \ifhmode
   \edef\@sf{\spacefactor=\the\spacefactor}\unskip${}#1$\relax\fi\fi}
\def\attach#1{\space@ver{\strut^{\mkern 2mu #1} }\@sf\ }
\newtoks\foottokens
%
%
%
%
\newbox\leftpage \newdimen\fullhsize \newdimen\hstitle \newdimen\hsbody
\newif\ifreduce  \reducefalse
\def\almostshipout#1{\if L\lr \count2=1
      \global\setbox\leftpage=#1 \global\let\lr=R
  \else \count2=2
    \shipout\vbox{\special{dvitops: landscape}
      \hbox to\fullhsize{\box\leftpage\hfil#1}} \global\let\lr=L\fi}
\def\smallsize{\relax
\font\eightrm=cmr8 \font\eightbf=cmbx8 \font\eighti=cmmi8
\font\eightsy=cmsy8 \font\eightsl=cmsl8 \font\eightit=cmti8
\font\eightt=cmtt8
\def\eightpoint{\relax
\textfont0=\eightrm  \scriptfont0=\sixrm
\scriptscriptfont0=\sixrm
\def\rm{\fam0 \eightrm \f@ntkey=0}\relax
\textfont1=\eighti  \scriptfont1=\sixi
\scriptscriptfont1=\sixi
\def\oldstyle{\fam1 \eighti \f@ntkey=1}\relax
\textfont2=\eightsy  \scriptfont2=\sixsy
\scriptscriptfont2=\sixsy
\textfont3=\tenex  \scriptfont3=\tenex
\scriptscriptfont3=\tenex
\def\it{\fam\itfam \eightit \f@ntkey=4 }\textfont\itfam=\eightit
\def\sl{\fam\slfam \eightsl \f@ntkey=5 }\textfont\slfam=\eightsl
\def\bf{\fam\bffam \eightbf \f@ntkey=6 }\textfont\bffam=\eightbf
\scriptfont\bffam=\sixbf   \scriptscriptfont\bffam=\sixbf
\def\tt{\fam\ttfam \eightt \f@ntkey=7 }
\def\caps{\fam\cpfam \tencp \f@ntkey=8 }\textfont\cpfam=\tencp
\setbox\strutbox=\hbox{\vrule height 7.35pt depth 3.02pt width\z@}
\samef@nt}
\def\Eightpoint{\eightpoint \relax
  \ifsingl@\subspaces@t2:5;\else\subspaces@t3:5;\fi
  \ifdoubl@ \multiply\baselineskip by 5
            \divide\baselineskip by 4\fi }
\parindent=16.67pt
\itemsize=25pt
\thinmuskip=2.5mu
\medmuskip=3.33mu plus 1.67mu minus 3.33mu
\thickmuskip=4.17mu plus 4.17mu
\def\thinspace{\kern .13889em }
\def\negthinspace{\kern-.13889em }
\def\enspace{\kern.416667em }
\def\enskip{\hskip.416667em\relax}
\def\quad{\hskip.83333em\relax}
\def\qquad{\hskip1.66667em\relax}
\def\crr{\cropen{8.3333pt}}
\foottokens={\Eightpoint\singlespace}
\def\papersize{\SIZE\OFFSET\skip\footins=\bigskipamount}
\def\SIZE{\hsize=11.8truecm\vsize=17.5truecm}
\def\OFFSET{\voffset=-1.3truecm\hoffset=  .14truecm}
\def\attach##1{\space@ver{\strut^{\mkern 1.6667mu ##1} }\@sf\ }
\def\PH@SR@V{\doubl@true\baselineskip=20.08pt plus .1667pt minus .0833pt
             \parskip = 2.5pt plus 1.6667pt minus .8333pt }
\def\author##1{\vskip\frontpageskip\titlestyle{\tencp ##1}\nobreak}
\def\address##1{\par\kern 4.16667pt\titlestyle{\tenpoint\it ##1}}
\def\andaddress{\par\kern 4.16667pt \centerline{\sl and} \address}
\def\abstract{\vskip\frontpageskip\centerline{\twelverm ABSTRACT}
              \vskip\headskip }
\def\cases##1{\left\{\,\vcenter{\Tenpoint\m@th
    \ialign{$####\hfil$&\quad####\hfil\crcr##1\crcr}}\right.}
\def\matrix##1{\,\vcenter{\Tenpoint\m@th
    \ialign{\hfil$####$\hfil&&\quad\hfil$####$\hfil\crcr
      \mathstrut\crcr\noalign{\kern-\baselineskip}
     ##1\crcr\mathstrut\crcr\noalign{\kern-\baselineskip}}}\,}
\Tenpoint
}
\def\Smallsize{\smallsize\reducetrue
\let\lr=L
\hstitle=8truein\hsbody=4.75truein\fullhsize=24.6truecm\hsize=\hsbody
\output={
  \almostshipout{\leftline{\vbox{\makeheadline
  \pagebody\makefootline}}}\advancepageno
     }
\special{dvitops: landscape}
\def\makeheadline{
\iffrontpage\line{\the\headline}
             \else\vskip .0truecm\line{\the\headline}\vskip .5truecm \fi}
\def\makefootline{\iffrontpage\vskip  0.truecm\line{\the\footline}
               \vskip -.15truecm\line{\the\date\hfil}
              \else\line{\the\footline}\fi}
\paperheadline={
\iffrontpage\hfil
               \else
               \tenrm\hss $-$\ \folio\ $-$\hss\fi    }
\paperstyle}
%
%
%
%
%
%
%
%
%
\newcount\referencecount     \referencecount=0
\newif\ifreferenceopen       \newwrite\referencewrite
\newtoks\rw@toks
\def\NPrefmark#1{\attach{\scriptscriptstyle [ #1 ] }}
\let\PRrefmark=\attach
\def\refmark#1{\relax\ifPhysRev\PRrefmark{#1}\else\NPrefmark{#1}\fi}
\def\refend{\refmark{\number\referencecount}}
\newcount\lastrefsbegincount \lastrefsbegincount=0
\def\refsend{\refmark{\count255=\referencecount
   \advance\count255 by-\lastrefsbegincount
   \ifcase\count255 \number\referencecount
   \or \number\lastrefsbegincount,\number\referencecount
   \else \number\lastrefsbegincount-\number\referencecount \fi}}
\def\refch@ck{\chardef\rw@write=\referencewrite
   \ifreferenceopen \else \referenceopentrue
   \immediate\openout\referencewrite=referenc.texauxil \fi}
%
{\catcode`\^^M=\active 
  \gdef\obeyendofline{\catcode`\^^M\active \let^^M\ }}%
%
{\catcode`\^^M=\active 
  \gdef\ignoreendofline{\catcode`\^^M=5}}
{\obeyendofline\gdef\rw@start#1{\def\t@st{#1} \ifx\t@st\blankend%
\endgroup \@sf \relax \else \ifx\t@st\bl@nkend \endgroup \@sf \relax%
\else \rw@begin#1
\backtotext
\fi \fi } }
{\obeyendofline\gdef\rw@begin#1
{\def\n@xt{#1}\rw@toks={#1}\relax%
\rw@next}}
\def\blankend{}
{\obeylines\gdef\bl@nkend{
}}
\newif\iffirstrefline  \firstreflinetrue
\def\rwr@teswitch{\ifx\n@xt\blankend \let\n@xt=\rw@begin %
 \else\iffirstrefline \global\firstreflinefalse%
\immediate\write\rw@write{\noexpand\obeyendofline \the\rw@toks}%
\let\n@xt=\rw@begin%
      \else\ifx\n@xt\rw@@d \def\n@xt{\immediate\write\rw@write{%
        \noexpand\ignoreendofline}\endgroup \@sf}%
             \else \immediate\write\rw@write{\the\rw@toks}%
             \let\n@xt=\rw@begin\fi\fi \fi}
\def\rw@next{\rwr@teswitch\n@xt}
\def\rw@@d{\backtotext} \let\rw@end=\relax
\let\backtotext=\relax

\newdimen\refindent     \refindent=30pt
\def\refitem#1{\par \hangafter=0 \hangindent=\refindent \Textindent{#1}}
\def\REFNUM#1{\space@ver{}\refch@ck \firstreflinetrue%
 \global\advance\referencecount by 1 \xdef#1{\the\referencecount}}
\def\refnum#1{\space@ver{}\refch@ck \firstreflinetrue%
 \global\advance\referencecount by 1 \xdef#1{\the\referencecount}\refend}

\def\REF#1{\REFNUM#1%
 \immediate\write\referencewrite{%
 \noexpand\refitem{#1.}}%
\begingroup\obeyendofline\rw@start}
\def\ref{\refnum\?%
 \immediate\write\referencewrite{\noexpand\refitem{\?.}}%
\begingroup\obeyendofline\rw@start}
\def\Ref#1{\refnum#1%
 \immediate\write\referencewrite{\noexpand\refitem{#1.}}%
\begingroup\obeyendofline\rw@start}
\def\REFS#1{\REFNUM#1\global\lastrefsbegincount=\referencecount
\immediate\write\referencewrite{\noexpand\refitem{#1.}}%
\begingroup\obeyendofline\rw@start}
\newif\ifmref  
\newif\iffref  
\def\xrefsend{\xrefmark{\count255=\referencecount
\advance\count255 by-\lastrefsbegincount
\ifcase\count255 \number\referencecount
\or \number\lastrefsbegincount,\number\referencecount
\else \number\lastrefsbegincount-\number\referencecount \fi}}
\def\xrefsdub{\xrefmark{\count255=\referencecount
\advance\count255 by-\lastrefsbegincount
\ifcase\count255 \number\referencecount
\or \number\lastrefsbegincount,\number\referencecount
\else \number\lastrefsbegincount,\number\referencecount \fi}}
\def\xREFNUM#1{\space@ver{}\refch@ck\firstreflinetrue%
\global\advance\referencecount by 1
\xdef#1{\xrefend}}
\def\xrefend{\xrefmark{\number\referencecount}}
\def\xrefmark#1{{#1}}
\def\xRef#1{\xREFNUM#1\immediate\write\referencewrite%
{\noexpand\refitem{#1 }}\begingroup\obeyendofline\rw@start}%
\def\xREFS#1{\xREFNUM#1\global\lastrefsbegincount=\referencecount%
\immediate\write\referencewrite{\noexpand\refitem{#1 }}%
\begingroup\obeyendofline\rw@start}
\referencecount=0
\def\refbreak{\hfil\penalty200\hfilneg}
\def\paperstyle{\papers}
\paperstyle   
%
%
%
\def\slacpub{\afterassignment\slacp@b\toks@}
\def\slacp@b{\edef\n@xt{\Pubnum={CERN--TH.\the\toks@}}\n@xt}
\let\pubnum=\slacpub
\expandafter\ifx\csname eightrm\endcsname\relax
    \let\eightrm=\ninerm \let\eightbf=\ninebf \fi

\font\seventeencp=cmcsc10 scaled\magstep3

\newif\ifCONF \CONFfalse
\newif\ifBREAK \BREAKfalse
\newif\ifsectionskip \sectionskiptrue

%
%
%
%
\def\NuclPhysProc{
\let\lr=L
\hstitle=8truein\hsbody=4.75truein\fullhsize=21.5truecm\hsize=\hsbody
\hstitle=8truein\hsbody=4.75truein\fullhsize=20.7truecm\hsize=\hsbody
\output={
  \almostshipout{\leftline{\vbox{\makeheadline
  \pagebody\makefootline}}}\advancepageno
     }
\def\papersize{\SIZE\OFFSET\skip\footins=\bigskipamount}
\def\SIZE{\hsize=10.0truecm\vsize=27.0truecm}
\def\OFFSET{\voffset=-1.4truecm\hoffset=-2.40truecm}
\def\makeheadline{
\iffrontpage\line{\the\headline}
             \else\vskip .0truecm\line{\the\headline}\vskip .5truecm \fi}
\def\makefootline{\iffrontpage\vskip  0.truecm\line{\the\footline}
               \vskip -.15truecm\line{\the\date\hfil}
              \else\line{\the\footline}\fi}
\paperheadline={\hfil}
\paperstyle}
%
%
%
%

\paperstyle
%
%
%
%
\def\ReprintVolume{\smallsize
\def\papersize{\hsize=18.0truecm\vsize=25.1truecm\voffset -.73truecm
    \hoffset -.65truecm\skip\footins=\bigskipamount
    \normaldisplayskip= 20pt plus 5pt minus 10pt}
\paperstyle\baselineskip=.425truecm\parskip=0truecm
\def\makeheadline{
\iffrontpage\line{\the\headline}
             \else\vskip .0truecm\line{\the\headline}\vskip .5truecm \fi}
\def\makefootline{\iffrontpage\vskip  0.truecm\line{\the\footline}
               \vskip -.15truecm\line{\the\date\hfil}
              \else\line{\the\footline}\fi}
\paperheadline={
\iffrontpage\hfil
               \else
               \tenrm\hss $-$\ \folio\ $-$\hss\fi    }
\def\sectionfont{\bf}    }
%
%
%
%
\def\SIZE{\hsize=15.73truecm\vsize=23.11truecm}
\def\OFFSET{\voffset=0.4truecm\hoffset=-0.88truecm}
\def\papersize{\SIZE\OFFSET\skip\footins=\bigskipamount
\normaldisplayskip= 30pt plus 5pt minus 10pt}
\def\CERN{\address{{\sl CERN, 1211 Geneva 23, Switzerland\
\phantom{XX}\ }}}
\Pubnum={\rm CERN$-$TH.\the\pubnum }
\def\title#1{\vskip\frontpageskip\vskip .50truein
     \titlestyle{\seventeencp #1} \vskip\headskip\vskip\frontpageskip
     \vskip .2truein}
\def\author#1{\vskip .27truein\titlestyle{#1}\nobreak}

\def\p@bblock{\begingroup \tabskip=\hsize minus \hsize
   \baselineskip=1.5\ht\strutbox \topspace-2\baselineskip
   \halign to\hsize{\strut ##\hfil\tabskip=0pt\crcr
   \the \Pubnum\cr}\endgroup}
\def\makefootline{\iffrontpage\vskip .27truein\line{\the\footline}
                 \vskip -.1truein\line{\the\date\hfil}
              \else\line{\the\footline}\fi}
\paperfootline={\iffrontpage
 \the\Pubnum\hfil\else\hfil\fi}
\paperheadline={
\iffrontpage\hfil
               \else
               \twelverm\hss $-$\ \folio\ $-$\hss\fi}
%
%
\def\nup#1({\refbreak\ Nucl.\ Phys.\ $\underline {B#1}$\ (}
\def\plt#1({\refbreak\ Phys.\ Lett.\ $\underline  {#1}$\ (}
\def\cmp#1({\refbreak\ Commun.\ Math.\ Phys.\ $\underline  {#1}$\ (}
\def\prp#1({\refbreak\ Physics\ Reports\ $\underline  {#1}$\ (}
\def\prl#1({\refbreak\ Phys.\ Rev.\ Lett.\ $\underline  {#1}$\ (}
\def\prv#1({\refbreak\ Phys.\ Rev. $\underline  {D#1}$\ (}
\def\und#1({            $\underline  {#1}$\ (}
%
%

%
%
\hyphenation{sym-met-ric anti-sym-me-tric re-pa-ra-me-tri-za-tion
Lo-rentz-ian a-no-ma-ly di-men-sio-nal two-di-men-sio-nal}
%
%
%
%

\def\coeff#1#2{{\textstyle { #1 \over #2}}\displaystyle}
\def\boxit#1{\vbox{\hrule\hbox{\vrule\kern3pt
\vbox{\kern3pt#1\kern3pt}\kern3pt\vrule}\hrule}}
\message{ by V.K, W.L and A.S}
\catcode`@=12
\paperstyle
\def\del{\partial}
\def\Fb{\overline{F}}
\def\nablab{\overline{\nabla}}
\def\Ub{\overline{U}}
\def\Db{\overline{D}}
\def\zb{\overline{z}}
\def\eb{\overline{e}}
\def\fb{\overline{f}}
\def\tb{\overline{t}}
\def\Xb{\overline{X}}
\def\Vb{\overline{V}}
\def\Cb{\overline{C}}
\def\Sb{\overline{S}}
\def\cy{Calabi-Yau}
\def\cabg{C_{\alpha\beta\gamma}}
\def\A{\Lambda}
\def\Ah{\hat \Lambda}
\def\B{\Sigma}
\def\Bh{\hat \Sigma}
\def\Kh{\hat{K}}
\def\Knh{{\cal K}}
\def\R{\hat{R}}
\def\V{{\cal V}}
\def\T{T}
\def\C{{\cal T}}
\def\Gammah{\hat{\Gamma}}
\def\Sigmat{\Sigma^{(3)}}
\def\twot{$(2,2)$}
\pubnum={6334/91}
\date{December 1991}
\titlepage
\title{Flat Holomorphic Connections and
Picard-Fuchs Identities From $N=2$ Supergravity}
\author{Sergio Ferrara and Jan Louis}
\CERN
\abstract
We show that in special K\"ahler geometry of $N=2$ space-time
supergravity the gauge variant part of the connection is holomorphic
and flat (in a Riemannian sense).
A set of differential identities (Picard-Fuchs identities)
are satisfied on a holomorphic bundle. The relationship
with the
differential equations obeyed by the periods of the holomorphic three
form of Calabi-Yau manifolds is outlined.
\endpage
\chapternumber=0
\pagenumber=2

\section{Introduction}
\REF\bd{For a review see L.~Dixon, in Proc. of the 1987 ICTP Summer
Workshop in High Energy Physics and Cosmology, Trieste,
 ed.~G.~Furlan,
J.~C.~Pati, D.~W.~Sciama, E.~Sezgin and Q.~Shafi and references therein.}
\REF\review{For a review see S.~Ferrara\journal Mod. Phys. Lett. & A6
 (91) 2175 and references therein.}
For the study of $N=1$
superstring vacua in four space-time dimensions one
is particularly interested in super conformal field theories
(SCFT) with central charge $c=9$ and
worldsheet supersymmetry $(0,2)$.\refmark{\bd}
For the subset of string vacua where
this worldsheet supersymmetry is enlarged to $(2,2)$ the low energy
effective Lagrangian satisfies an additional constraint.
The manifold spanned by certain massless scalar fields in the string
spectrum (so called moduli)
is promoted from being a
K\"ahler manifold to a
`special K\"ahler manifold'.\refmark{\review}
\REF\pw{B.~de Wit and A.~van Proeyen\journal Nucl. Phys. & B245 (84) 89;
\brk B.~de Wit, P.~Lauwers and A.~van Proeyen\journal
Nucl. Phys. & B255 (85) 569.}
\REF\ckvdfdg{E.~Cremmer, C.~Kounnas, A.~van Proeyen,{}
 J.~P.~Derendinger, S.~Ferrara, B.~de Wit and L.~Girardello\journal
Nucl. Phys.& B250 (85) 385.}
Such a geometric structure first appeared
in the context of
$N=2$ supergravity theories in four space-time dimensions
as the scalar manifold spanned by the scalars of
vectormultiplets.\refmark{\pw,\ckvdfdg}
\REF\ns{N.~Seiberg\journal Nucl. Phys.& B303 (88) 206.}
\REF\cfg{S.~Cecotti, S.~Ferrara and L.~Girardello\journal
 Int. J. Mod. Phys.& 4 (89) 2475\journal Phys. Lett. & B213 (88) 443.}
\REF\dkl{L.~Dixon, V.~Kaplunovsky and J.~Louis\journal
Nucl. Phys.&B329  (90) 27}
The moduli space of \twot\  SCFT  with central charge $c=9$
obeys a similar constraint as can be understood from
the relation between heterotic and type II string
theories\refmark{\ns,\cfg}
or directly from SCFT Ward identities.\refmark{\dkl}
\REF\fs{S.~Ferrara and A.~Strominger, in {\it Strings 89}, eds.
        R.~Arnowitt, R.~Bryan, M.~Duff, D.~Nanopoulos and C.~Pope,
        World Scientific, Singapore,1989.}
\REF\as{A.~Strominger\journal Comm. Math. Phys.& 133 (90) 261.}
\REF\co{P.~Candelas and X.~de la Ossa\journal Nucl. Phys.
& B342 (90) 246.}
Calabi-Yau threefolds are an example of \twot\ SCFT and thus
their moduli spaces are also special K\"ahler manifolds.
This can also be derived by purely geometrical methods without
ever referring to a SCFT.\refmark{\fs,\as,\co}

\REF\cdgp{P.~Candelas, X.~de la Ossa, P.~Green and L.~Parkes\journal
Phys. Lett.&B258 (91) 118\journal
Nucl. Phys. & B359 (91) 21.}
\REF\gp{B.~Greene and M.~Plesser\journal Nucl. Phys.& B338 (90)
15.}
Recently, for a specific \cy\ manifold (a quintic in $CP_4$)
the exact K\"ahler and super-potential of the
low energy theory was given without using the underlying
SCFT.\refmark{\cdgp}
This was possible by explicitly constructing the mirror map\refmark{\gp}
 of this
\cy\ moduli space.
In the course of this investigation
it was shown that the periods of the holomorphic threeform $\Omega$
satisfy a linear differential equation.
\REF\cf{A.~Cadavid and S.~Ferrara\journal Phys. Lett.& B267 (91) 193.}
\REF\lsw{W.~Lerche, D.~J.~Smit and N.~Warner,{}
Caltech preprint CALT-68-1738.}
\REF\dm{D.~Morrison, Duke preprint DUK-M-91-14.}
It was then realized that this equation is
 a particular case of Picard-Fuchs
systems of differential equations, known to be obeyed by the periods
of globally defined $p$-forms ($p=3$ for a threefold) from algebraic
geometry.\refmark{\cf-\dm}
\REF\dvvtop{R.~Dijkgraaf, E.~Verlinde and H.~Verlinde\journal Nucl.
  Phys. & B352 (91) 59.}
\REF\bv{B.~Blok and A.~Varchenko, Princeton preprint IASSNS-HEP-91/5.}
\REF\cgp{S.~Cecotti, S.~Girardello and A.~Pasquinucci\journal Nucl. Phys.
&B328 (89) 701,\brk
 Int. J. Mod. Phys. {\bf A6} (1991) 2427.}
\REF\cv{S.~Cecotti and C.~Vafa, Harvard preprint HUTP-91/A031.}
\REF\vw{E.~Verlinde and N.~Warner\journal Phys. Lett. &B269 (91) 96.}
\REF\zm{Z.~Maassarani, USC preprint USC-91/023.}
\REF\kst{A.~Klemm, M.~G.~Schmidt and S.~Theisen,{}
Karlsruhe preprint KA-THEP-91-09.}
Furthermore,
substantial information about
$(2,2)$ SCFT can be obtained from a purely topological
sector of the theory.\refmark{\dvvtop-\cv}
For $c=3$ SCFT analogous differential equations were derived
using methods developed in
topological field theories (TFT).\refmark{\vw-\kst}

Since the moduli space of \cy-threefolds or more generally the moduli
space of $(2,2)$ SCFT of central charge $c=9$
is constrained to display special geometry
it should be possible to understand the Picard-Fuchs equations
entirely within the framework of special geometry. This is
the aim of this paper.
We first briefly remind the reader of the properties
of special K\"ahler geometry.
Then we show that the Christoffel as well as
the K\"ahler connection consist of two
distinctively different pieces. A holomorphic part is responsible
for the transformation properties of the connection whereas the
non-holomorphic piece transforms like a tensor.
The holomorphic part of the connection turns out to be flat
(in a Riemannian sense).
We believe that this holomorphic flat connection
is the analogue of the flat  connection encountered in
TFT when it is restricted to marginal deformations.\refmark{\bv,\cv,\lsw}

Equipped with this understanding
we then show how special geometry implies a set of covariant
 differential identities (which we call Picard-Fuchs identities).
These identities are entirely equivalent to the identity the
Riemann tensor satisfies in special geometry
but are instead of purely holomorphic  nature.
Once the coefficient functions
of these identities are specified a non-trivial linear
differential equation
arises which is equivalent to the Picard-Fuchs equation satisfied
by the periods of the \cy-manifold.

%
\section{Summary of special geometry}
The metric $g_{\alpha \bar \beta}, \alpha = 1,\ldots,n$ of a
$n$-dimensional K\"ahler
manifold is given by the second derivative of a K\"ahler potential $K$ :
$g_{\alpha \bar \beta} = \del_\alpha \bar \del_{\bar \beta} K
\, (\equiv K_{\alpha \bar \beta})$.
$K$ is a real function of the complex coordinates $z^\alpha$ and
$\zb^{\bar \alpha}$.
$g_{\alpha \bar \beta}$
is invariant under the K\"ahler
transformations
$K(z,\zb) \rightarrow  K(z,\zb) + f(z) + \fb (\zb)$.
A field $\Psi$
of K\"ahler weight
$(k,\bar k)$ transforms according to
$\Psi \rightarrow \Psi e^{kf} e^{\bar k \fb}$.
Its K\"ahler covariant derivative is defined by
$$
\eqalign{
D_\alpha \Psi  = (\nabla_\alpha - k K_\alpha) \Psi \cr
\Db_{\bar \alpha} \Psi  =
(\nablab_{\bar \alpha} - \bar k K_{\bar \alpha})
\Psi }
\eqn\kcovd
$$
where $\nabla_\alpha$ denotes the covariant derivative with respect to
reparametrisations of the K\"ahler manifold.
Thus $K_\alpha, K_{\bar \alpha}$ act as connections for K\"ahler
transformations.
This is a consequence of the fact that special manifolds are manifolds
of restricted type. This means that $K_\alpha$ is an abelian
 connection
of a holomorphic line bundle $L$ whose first Chern class is the
K\"ahler class ($c_1(L) = [J]$).\refmark{\review}

Special geometry is defined by an additional constraint on the
K\"ahler metric.
This constraint can be expressed as a covariant equation the
Riemann tensor of the K\"ahler manifold satisfies.
It reads\refmark{\ckvdfdg}
$$
R_{\alpha \bar \beta \gamma}^{ \delta} =
g_{\alpha \bar \beta} \delta_{\gamma}^{\delta} +
g_{\gamma \bar \beta} \delta_\alpha^\delta     -
C_{\alpha\gamma\epsilon} g^{\epsilon \bar \epsilon}
\Cb_{\bar \beta \bar \delta \bar \epsilon}  g^{\delta\bar \delta}
\eqn\rcrelation
$$
where $C_{\alpha\beta\gamma}$ is of K\"ahler weight $(-1,1)$,
symmetric in its indices and satisfies
$$
\eqalign{
\Db_{\bar \epsilon} C_{\alpha\beta\gamma} =& 0\, ,\cr
D_{\epsilon} C_{\alpha\beta\gamma} -
D_{\alpha} C_{\epsilon\beta\gamma} =&0 \, .}
\eqn\dc
$$
Using $D_\alpha e^K = 0$
eq.~\dc\ can be solved by
$$
\eqalign{
C_{\alpha\beta\gamma} =&
e^K W_{\alpha\beta\gamma}(z) \, ,\cr
W_{\alpha\beta\gamma} =&
D_\alpha D_\beta D_\gamma S(z,\zb) \, ,\cr
\bar \del_{\bar \epsilon} W_{\alpha\beta\gamma} =&0 \, . }
\eqn\cwrelation
$$
\REF\cdf{L.~Castellani, R.~D'Auria and S.~Ferrara\journal Phys. Lett&
         B241 (90) 57\journal Class. Quant. Grav. &1 (90) 317;\brk
R.~D'Auria, S.~Ferrara and P.~Fr\'e\journal Nucl. Phys. & B359 (91) 705.}
Also, one can find $K$ and $S$ such that
eq.~\rcrelation\ is satisfied in arbitrary coordinates.\refmark{\cdf}
This is achieved by
introducing $n+1$ holomorphic sections $X^A(z), A = 0,1,\ldots,n$
which obey
$\bar{\del}_{\bar \alpha} X^A = 0$. In terms of $X^A$ the K\"ahler
potential reads
$$
\eqalign{
K&= - \ln Y \cr
Y&= X^A N_{AB} \Xb^{B} = X^A \Fb_{A} + \Xb^A F{_A}    }
\eqn\ksol
$$
where
$$
\eqalign{
N_{AB}(X,\Xb) =& F_{AB}(X) + \Fb_{AB}(\Xb)  \cr
F_{AB}(X) =&\del_A \del_B F(X)  }
\eqn\nfdef
$$
and $F(X)$ is a homogeneous function of degree two.
$C_{\alpha\beta\gamma}$ and $S$ are then given by
$$
\eqalign{
W_{\alpha\beta\gamma} =&  \del_\alpha X^A \del_\beta X^B
\del_\gamma X^C F_{ABC} \cr
S =&- {1 \over 2}  X^A N_{AB} X^{B} \, .}
\eqn\sxrelation
$$
$X^A$ transforms under K\"ahler transformations with weight $(-1,0)$,
$\Xb^A$ with weight $(0,-1)$. $N_{AB}$ is invariant as a consequence
of $F$ being homogeneous of degree two.

Let us define the ($2(n+1)$ dimensional symplectic) holomorphic
vector $V(z) = (X^A (z), iF_A (z))$.
It is straightforward to show that as a consequence of
eqs.~\ksol-\sxrelation\
$V(z)$ satisfies  the following set
of identities (and its complex conjugate)\refmark{\cdf}
$$
\eqalign{
D_\alpha V &= U_\alpha \cr
D_\alpha U_\beta &= C_{\alpha\beta\gamma} g^{\gamma \bar \gamma}
\Ub_{\bar \gamma} \cr
D_\alpha \Ub_{\bar \beta} &= g_{\alpha \bar \beta} \Vb \cr
D_\alpha \Vb &= 0                                 }
\eqn\deqset
$$
where the first equation defines $U_\alpha$.\foot{
The exact same set of identities were derived in a \cy\ context in
ref.~\as,\co.}
Eq.~\rcrelation\  now
follows from eqs.~\deqset.
Thus \deqset\ is an alternative way to
 characterize special geometry.

Finally, there is a particularly simple set of coordinates given
by
$
z^\alpha= X^a/X^0, a=1,\ldots,n
$.\refmark{\pw,\cdf}
In these so called `special coordinates' $K$ and
$C_{\alpha\beta\gamma}$ reduce to\foot{
With a further (K\"ahler) gauge choice one can set $X^0 =1$. However,
special coordinates do not require this gauge.}
$$
\eqalign{
K =& - \ln \left[ 2(F+\Fb) - (F_{\alpha} - \Fb_{\bar \alpha})(z^\alpha -
\zb^{\bar \alpha}) \right] \, \cr
C_{\alpha\beta\gamma} =& Y^{-1}  F_{\alpha\beta\gamma}  \, .}
\eqn\kcspcoo
$$
We will see in the next section that these coordinates are
the flat coordinates for a holomorphic connection.

\REF\lvw{W.~Lerche, C.~Vafa and N.~Warner\journal Nucl. Phys.&B324 (89)
427.}
As we remarked in the introduction
special K\"ahler geometry
arises in $N=2$ supergravity theories in four space-time dimensions
as the manifold spanned by the scalars in the
vectormultiplets.\refmark{\pw,\ckvdfdg}
The moduli space of \twot\ SCFT (and thus the moduli space of
 \cy-threefolds)
also obeys eq.~\rcrelation\ where
$C_{\alpha\beta\gamma}$ are the Yukawa couplings
(the structure constants of the chiral ring\refmark{\lvw})
of the associated
matter multiplets.\refmark{\dkl}

\section{holomorphic connection}
Before we turn to the study of eqs.~\deqset\ let us first make a few
observations about the connections
in special geometry. We already remarked that in addition to
the usual  Christoffel connection
$\Gamma_{\alpha\beta}^\gamma$ there also is a K\"ahler connection
given by $K_\alpha$. However, due to the additional constraint
on the K\"ahler potential (eqs.~\rcrelation,\deqset) we find that
both connections enjoy further properties in special geometry.
These are most easily displayed by introducing
 K\"ahler invariant functions $t^a$ as follows
$$
t^a(z) = { X^a(z) \over X^0(z)} ,
\qquad  a=1,\ldots,n \, .
\eqn\tdef
$$
In terms of $t^a$ and $X^0$  the K\"ahler connection $K_\alpha$
decomposes   into a sum of
a purely holomorphic term and a non-holomorphic piece
$$
K_\alpha (z,\zb) =
\Kh_\alpha (z) +
\Knh_\alpha (z,\zb)
\eqn\kconndef
$$
where
$$
\eqalign{
\Knh_\alpha (z,\zb) =&e_\alpha^a(z) K_a(z,\zb) \equiv
e_\alpha^a(z)  {\del\over\del t^a} K(t(z),\tb(\zb))          \cr
\Kh_\alpha (z) =& - \del_\alpha \ln X^0 (z) \cr
e_\alpha^a(z) =&\del_\alpha t^a (z) }
\eqn\hconndef
$$
The significance of the holomorphic piece $\Kh_\alpha$ is that it
transforms as a connections whereas $\Knh_\alpha$ is invariant
under K\"ahler transformation:
$$
K_\alpha \rightarrow K_\alpha + f_\alpha\, , \qquad
\Kh_\alpha \rightarrow \Kh_\alpha + f_\alpha\, ,   \qquad
\Knh_\alpha \rightarrow \Knh_\alpha     \, .
\eqn\ktrans
$$

A very similar structure also occurs for
$\Gamma_{\alpha\beta}^\gamma$. Let us first compute the
metric in terms of $t^a$ and $X^0$:
$$
g_{\alpha\bar \beta} = e_\alpha^a \eb_{\bar \beta}^{\bar b} g_{a\bar b}
\equiv e_\alpha^a \eb_{\bar \beta}^{\bar b}
{\del \over \del t^a}
{\del \over \del \bar{t}^{\bar b}} K (z,\zb)  \, .
\eqn\metricdef
$$
(We see that
$e_\alpha^a(z)$ can be viewed as a holomorphic vielbein.)
As a consequence
the Christoffel connection ($\Gamma_{\alpha\beta}^\gamma \equiv
g_{\alpha\bar \gamma \beta} g^{-1 \bar \gamma \gamma}$)
also splits in the following way
$$
\Gamma_{\alpha\beta}^\gamma (z,\zb) =
\Gammah_{\alpha\beta}^\gamma (z) +
\T_{\alpha\beta}^\gamma (z,\zb)
\eqn\conndef
$$
where
$$
\eqalign{
\T_{\alpha\beta}^\gamma (z,\zb) = &
e_{\alpha}^a e_\beta^b
g_{a \bar d b} g^{-1 \bar d c} e_c^{-1 \gamma}     \cr
\Gammah_{\alpha\beta}^\gamma (z) =& (\del_\beta e_\alpha^a)
 e^{-1 \gamma}_a }
\eqn\tconndef
$$
Again the holomorphic
$\Gammah_{\alpha\beta}^\gamma (z)$ transforms as a connection
under reparametrisations
whereas
$\T_{\alpha\beta}^\gamma$
enjoys tensorial transformation properties.
Thus in both cases the
transformation properties of the connections
are entirely carried by the holomorphic objects
$\Kh$ and $\Gammah$. Thus it is possible to introduce a purely
holomorphic covariant derivative $\hat D$ where $\Gamma$ and $K$
are replaced by $\Gammah$ and $\Kh$.
As we will see in the next section the
holomorphic (Picard-Fuchs) identities
precisely use $\hat D$.

The (holomorphic) metric for which
$\Gammah$ is the connection reads
$$
\hat{g}_{\alpha\beta} = e_\alpha^a e_\beta^b \eta_{ab}
\eqn\etadef
$$
where $\eta_{ab}$ is a constant (invertible) symmetric matrix.
Note that
$\hat{g}_{\alpha\beta}$
has two holomorphic indices in contrast to the K\"ahler metric
$g_{\alpha\bar\beta}$.
{}From eq.~\tconndef\ and \etadef\ it comes as no surprise
that $\Gammah$ also satisfies
$$
\R_{\delta \alpha\beta}^\gamma \equiv
\del_\delta \Gammah_{\alpha\beta}^\gamma
- \del_\alpha \Gammah_{\delta\beta}^\gamma
+ \Gammah_{\alpha\beta}^\mu  \Gammah_{\mu\delta}^\gamma
- \Gammah_{\delta\beta}^\mu  \Gammah_{\mu\alpha}^\gamma =0
\eqn\flatconn
$$
which means that it is a flat Riemannian connection.
The flat coordinates are exactly
the special
coordinates ($t^a = z^\alpha$) introduced in the last section.
In these coordinates we find
$$
e_\alpha^a = \delta_\alpha^a\, ,\qquad
\Gammah_{\alpha\beta}^\gamma = 0 \, ,  \qquad
\hat{g}_{\alpha\beta} =\eta_{\alpha\beta}  \, .
\eqn\kgsc
$$
(The gauge choice $X^0=1$ then implies $\Kh_\alpha =0$.)

Before we turn to the derivation of the differential equation
let us collect a few more formulas which we use in the next section.
We observe that
eq.~\rcrelation\ can also be expressed as
$$
\del_{\bar \beta} \left( \Gamma_{\alpha \gamma}^{ \delta} -
K_{\alpha } \delta_{\gamma}^{\delta} -
K_{\gamma } \delta_\alpha^\delta     +
\C_{\alpha\gamma}^{\delta} \right) = 0
\eqn\rcrelationp
$$
where
$$
\C_{\alpha\gamma}^{\delta}  =
C_{\alpha\gamma\epsilon} g^{\epsilon \bar \epsilon}
(\Db_{\bar \delta} \Db_{\bar \epsilon} \Sb) g^{\delta\bar \delta}
= Y^{-1} D_\alpha X^A D_\gamma X^C
\Db_{\bar \beta} \Xb^B F_{ABC} g^{\bar\beta \delta} \, .
\eqn\cdef
$$
Inserting eqs.~\conndef\ and \kconndef\ into eq.~\rcrelationp\ we
learn that
$\T_{\alpha\gamma}^{\delta}$ can be further decomposed into
$$
\T_{\alpha \gamma}^{ \delta} =
\Knh_{\alpha } \delta_{\gamma}^{\delta} +
\Knh_{\gamma } \delta_\alpha^\delta  -
\C_{\alpha\gamma}^{\delta}
\eqn\tsplit
$$
Of course, this can also be verified directly from the definitions
of $\T_{\alpha\gamma}^\delta$ and $\Knh_\alpha$.

Another flat (non-holomorphic) connection exists on
 the flat symplectic bundle of ref.~\as.
What we find here is that a flat holomorphic connection exists on a
$n$-dimensional space rather than a $2n+2$ dimensional space as in
ref.~\as.
In the deformation theory of the Hodge structure of \cy-threefolds
the components of
the symplectic vector $V$ are precisely the periods of the holomorphic
three form $\Omega$.
We believe that $\Gammah$ is the analogue of the flat
connection of $N=2$ TFT when it is restricted to the marginal
deformations.\refmark{\bv,\cv,\lsw}
(In general the connection of the TFT acts in the space
of all topological deformations not only the marginal ones.)

\section{Holomorphic covariant Picard-Fuchs identities}
\REF\copc{P.~Candelas and X.~de la Ossa, private communication.}
Let us come back to the set of identities given in eq.~\deqset.
As we remarked above they are equivalent to the identity \rcrelation\
the Riemann tensor of special geometry satisfies.
The Riemann tensor
 identity turns into a non-trivial differential equation for
the K\"ahler potential $K$ if the Yukawa couplings $\cabg$ are
specified. In the same spirit the identities \deqset\
turn into a non-trivial differential equation once the coefficient
functions are specified. As we will see in a moment the difference
is that the latter leads to holomorphic differential equations
which are exactly the Picard-Fuchs equations
of the \cy-manifold\refmark{\copc,\cdgp,\cf-\dm}
 and the analogous equations
in TFT.\refmark{\vw-\kst}

First we consider the case where $\cabg=0$ holds.\foot{
This is not an unrealistic case but does occur on certain subspaces
of the moduli space of \twot\ SCFT.\refmark{\dkl} }
Then eqs.~\deqset\ reduce to
$$
D_\alpha D_\beta V(z)
 = (\del_\alpha \del_\beta + \A_{\alpha\beta}^\gamma
\del_\gamma + \B_{\alpha\beta}) V(z) = 0
\eqn\deqcz
$$
where
$$
\eqalign{
 \A_{\alpha\beta}^\gamma =& K_\alpha \delta_\beta^\gamma +
K_\beta \delta_\alpha^\gamma - \Gamma_{\alpha\beta}^\gamma \, , \cr
\B_{\alpha\beta} =&
K_{\alpha\beta} + K_\alpha K_\beta
 - \Gamma_{\alpha\beta}^\gamma K_\gamma \, . }
\eqn\abdef
$$

As we just discussed
eq.\deqcz\ is an identity if we use the explicit form
of $K$ given by eqs.~\ksol-\sxrelation.
However, the coefficients
$\A_{\alpha\beta}^\gamma$ and $\B_{\alpha\beta}$
satisfy two noteworthy features. First, they transform
under coordinate and K\"ahler transformations as to render eq.~\deqcz\
covariant. This is obvious from eq.~\deqcz\ but can also be checked
explicitly in eq.~\abdef. Secondly, they
satisfy
$
\bar \del_{\bar \delta} \A_{\alpha\beta}^\gamma =
\bar \del_{\bar \delta} \B_{\alpha\beta} = 0
$
which holds as a consequence of eq.~\rcrelation. Again, this
has to be the case since $V(z)$ is  holomorphic
 by definition.
The considerations of the last section explain
how it is possible to have manifestly holomorphic $\A,\B$ which at the
same time transform appropriately: they can only depend on the
holomorphic connections $\Kh$ and $\Gammah$.
Indeed, inserting eqs.~\kconndef\ and \conndef\ into
\abdef\ one verifies
$$
\eqalign{
 \A_{\alpha\beta}^\gamma = &
 \Ah_{\alpha\beta}^\gamma \equiv \Kh_\alpha \delta_\beta^\gamma +
\Kh_\beta \delta_\alpha^\gamma - \Gammah_{\alpha\beta}^\gamma \, ,\cr
\B_{\alpha\beta} =  &
\Bh_{\alpha\beta} \equiv
\Kh_{\alpha\beta} + \Kh_\alpha \Kh_\beta
 - \Gammah_{\alpha\beta}^\gamma \Kh_\gamma \, . }
\eqn\abholo
$$
Thus eq.~\deqcz\ is covariant and holomorphic and the coefficients
$\A$ and $\B$ depend on the holomorphic connections defined in the
last section. Thus, if these holomorphic coefficients are given
eq.~\deqcz\ turns into a non-trivial differential equation for $V$.
Of course in this simple case the solution is already known.
In special coordinates  (and in the gauge $X^0=1$)
eq.~\deqcz\ simplifies further and reads
$
\del_\alpha \del_\beta X^A =
\del_\alpha \del_\beta F_A = 0
$
which is consistent with our input
$\cabg = F_{\alpha\beta\gamma} = 0$.
The solutions
 are (at least locally) the homogeneous spaces
$SU(1,n)/SU(n)\times U(1)$\foot{
Globally, there can be discrete identifications which leads
to the quotient of the homogeneous space with the duality group.
The latter is related to the monodromy group of
 eq.~\deqcz.\refmark{\cf,\lsw}  }
 with
$
F = c_{AB} X^A X^B
$
\REF\cp{E.~Cremmer and A.~van Proeyen\journal Class. Quant. Grav.&
2 (85) 445.}
where $c_{AB}$ is some constant symmetric matrix.\refmark{\pw,\cp}

Let us turn to the case of non vanishing $\cabg$.
Now  $D_\alpha D_\beta V$ is non-zero and $\A,\B$ are no longer
holomorphic. Instead they satisfy
$$
\eqalign{
 \A_{\alpha\beta}^\gamma = &
 \Ah_{\alpha\beta}^\gamma + \C_{\alpha \beta}^\gamma \cr
\B_{\alpha\beta} = &
\Bh_{\alpha\beta} +
\C_{\alpha\beta}^\gamma  K_\gamma
+  \Sigmat_{\alpha\beta}                    }
\eqn\abnonholo
$$
where
$$
\bar \del_{\bar \gamma} \Sigmat_{\alpha\beta} = -
\C_{\alpha\beta}^\gamma  g_{\gamma \bar\gamma}   \, .
\eqn\sigmthree
$$
Now a holomorphic identity appears if one uses the full set \deqset.
This is particularly simple if one considers a
one dimensional manifold.
In this case eqs.~\deqset\ are equivalent to
$$
DDW^{-1}DD V = \sum_{n=0}^4 a_n(z) \, \del^n V = 0  \, .
\eqn\oddeq
$$
The coefficients $a_n$ are given by\foot{
The dependence of $a_3$ on $W$ was also found in ref.\cdgp.}
$$
\eqalign{
a_4 = & W^{-1}\cr
a_3 = & 2 \del W^{-1}\cr
a_2 = & W^{-1} (\del \A - \A^2 + 2\B) + \del W^{-1} \A +
\del^2 W^{-1} \cr
a_1 = & W^{-1}(\del^2 \A + 2 \del \B - 2 \A \del \A)
      + \del W^{-1} (2\B + 2\del \A - \A^2)
      + \del^2 W^{-1} \A  \cr
a_0 = & W^{-1} (\B^2 - \B\del \A - \A \del \B + \del^2 \B)
      + \del W^{-1} (2 \del \B - \A \B)
      +  \del^2 W^{-1} \B  \cr    }
\eqn\acoeff
$$
where
$$
\eqalign{
\A =& 2\del K - \Gamma \cr
\B =& \del^2 K + (\del K)^2 - \Gamma \del K \, .}
\eqn\aboddef
$$
As before the
$a_n$'s have the appropriate transformation properties and satisfy
$
\bar \del a_n = 0
$
as a consequence of eq.~\rcrelation.
Inserting eq.~\conndef\ and \kconndef\ into \acoeff\ we find
that in all coefficients $a_n$ $K$ and $\Gamma$
are replaced by there holomorphic pieces $\Kh$ and $\Gammah$.
Thus in eq.~\acoeff\ instead of $\A$ and $\B$ appear
$$
\eqalign{
\Ah =& 2\del \Kh - \Gammah \, , \cr
\Bh =& \del^2 \Kh + (\del \Kh)^2 - \Gammah \del \Kh \, .}
\eqn\aboddef
$$
Now the meaning of eq.~\oddeq\ becomes transparent. As it stands
it is a covariant identity in special geometry. The coefficient functions
$a_n$ are holomorphic functions of the Yukawa couplings $W$ and
the holomorphic connections $\Kh,\Gammah$
(or equivalently the einbein $e$ and $X^0$).
If these three quantities
are given as `data'
eq.~\oddeq\ turns into a linear differential equation for $V$.
(Given $a_n$ eqs.~\acoeff\ also imply a non-linear differential
equation for $e$ and $X^0$.)

{}From eq.~\acoeff\ we learn
that the $a_n$ are
related in the following form
$$
a_3 = 2 \del a_4\, , \qquad
a_1 = \del a_2 - \coeff12 \del^2 a_3  \, .
\eqn\arelation
$$
Indeed, for the specific example of the quintic in $CP_4$
the $a_n$'s  (in the Landau-Ginzburg coordinates)
are given by\refmark{\cdgp,\cf,\lsw}
$$
\eqalign{
a_4 =  1- \psi^5 & \qquad a_3 =  - 10 \psi^4\qquad  a_2 = - 25 \psi^3 \cr
a_1 = & - 15 \psi^2\qquad a_0 =  - \psi }
\eqn\qcoeff
$$
and they obey \arelation.

In special coordinates (with the additional gauge choice $X^0=1$)
eq.~\oddeq\ takes a particularly simple form,
its coefficients read
$$
\eqalign{
a_4 = & W^{-1}\, , \qquad a_3 =  2 \del W^{-1}\, ,\cr
a_2 = & \del W^{-2}\, , \qquad a_1 =  a_0 = 0 \, . }
\eqn\acoeffsc
$$
The reason for the vanishing of $a_1$ and $a_0$ is that in these
coordinates
$\del X^0= 0$ and $\del X^1 =1$
holds. Thus for consistency  $a_0=a_1=0$ is required.
(Also, we
can use \acoeffsc\ and $W= \del^3 F$, $V= (1,X^1, i(2F - X^1 \del F),
i \del F)$ to
verify that \oddeq\ is  identically satisfied.)

Finally, it is possible to remove the K\"ahler connection $\Kh$
from eq.~\oddeq\ altogether. If we rescale $V$ by $1/X^0$ and define
a K\"ahler invariant $\V = V/X^0$ the differential equation reads
$$
\sum_{n=0}^4 a'_n(z) \, \del^n \V = 0  \, .
\eqn\oddeqp
$$
In this basis
$a'_n$ is independent of $\Kh$ and all
$X^0$ dependence appears through the ratio
$W/(X^0)^2$. In this basis $a_0=0$ holds which again is required for
consistency. Inverting this procedure produces a relation between
$a_0$ and
$a_1,a_2,a_3$.
\section{Conclusion}

In this letter we found that the connections
in special geometry can be split into a holomorphic piece
which transforms as a connection and a non-holomorphic piece with
tensorial transformation properties.
This fact allows for the existence of holomorphic covariant
identities in special geometry.
When certain `data' (the Yukawa coupling and the holomorphic
connection) are given these identities turn into linear differential
equations which are the exact analogue of the
Picard-Fuchs equations for the periods of the holomorphic threeform
$\Omega$ in \cy\ manifolds.

We treated in some detail the case of a one dimensional K\"ahler
manifold but clearly our considerations can be extended to
arbitrary dimension. From eq.~\deqset\
it is obvious that again holomorphic
identities occur whose coefficient function depend on the
holomorphic connections.

Our method is particularly powerful when applied to deformation theory
of \cy\ threefolds and $N=2$ TFT. Indeed, it allows to obtain
Picard-Fuchs equations in arbitrary coordinate systems and to give
a precise relation between the K\"ahler geometry of the moduli space
and the flat geometry of the deformation of the chiral ring restricted
to marginal operators. Various issues, discussed recently in the
literature can be further investigated and better understood such as
relation of Picard-Fuchs systems to $W$-algebras and target space
duality symmetry.

\ack

It is a pleasure to thank R.~D'Auria, A.Ceresole, M.~Cveti\v c,
W.~Lerche,
M.~Peskin, R.~Schimmrigk, D.~Smit and N.~Warner
 for useful discussions and in
particular P.~Candelas and X.~de la Ossa for communicating
some unpublished work. J.L. thanks the Aspen Center of Physics
for providing a stimulating atmosphere where this investigation started.

Some of the computations were performed using Maple.

\par \penalty-4000\vskip\chapterskip
   \spacecheck\referenceminspace \immediate\closeout\referencewrite
   \referenceopenfalse
   \line{\fourteenrm\hfil REFERENCES\hfil}\vskip\headskip
   \endlinechar=-1
   \input referenc.texauxil
   \endlinechar=13
   
\bye